\documentclass[prb,showpacs,preprintnumbers,amsmath,amssymb,printfigures,twocolumn,superscriptaddress,10pt]{revtex4-1}
\usepackage{longtable}
\usepackage[colorlinks=true,allcolors=blue]{hyperref}
\usepackage{graphicx}
\usepackage{amsmath}
\usepackage{bm}
\usepackage[T1]{fontenc}
\usepackage[latin9]{inputenc}
\usepackage{textcomp}
\usepackage{gensymb}
\usepackage{amstext}
\usepackage[Symbol]{upgreek}
\usepackage{booktabs}
\usepackage{dcolumn}
\usepackage{color}
\usepackage{enumerate}
\usepackage{latexsym,bm}
\usepackage{setspace}
\usepackage{braket}
\usepackage{float}
\usepackage{array}
\usepackage{multirow}
\makeatletter

\begin{document}
\title{Computational Search for Two-Dimensional Photocatalysts}
\author{V. Wang}
\thanks{wangvei@icloud.com}
\affiliation{Department of Applied Physics, Xi'an University of Technology, Xi'an 710054, China}  

\author{G. Tang}
\affiliation{Advanced Research Institute of Multidisciplinary Science, Beijing Institute of Technology, Beijing 100081, China} 

\author{Y. C. Liu}
\affiliation{Department of Applied Physics, Xi'an University of Technology, Xi'an 710054, China} 

\author{Y. Y. Liang}
\affiliation{Department of Physics, Shanghai Normal University, Shanghai 200234, China}  

\author{H. Mizuseki}
\affiliation{Korea Institute of Science and Technology (KIST), Seoul 02792, Republic of Korea}  

\author{Y. Kawazoe}
\affiliation{New Industry Creation Hatchery Center, Tohoku University, Sendai, Miyagi 980-8579, Japan} 
\affiliation{Department of Physics and Nanotechnology, SRM Institute of Science and Technology, Kattankulathur,Tamil Nadu-603203, India} 
\affiliation{Department of Physics, Suranaree University of Technology, Nakhon, Ratchasima, Thailand} 

\author{J. Nara}
\affiliation{National Institute for Materials Science, Tsukuba 305-0044, Japan}

\author{W. T. Geng}
\thanks{geng@hainanu.edu.cn}
\affiliation{School of Materials Science and Engineering, Hainan University, Haikou 570228, China}

\date{\today}

\begin{abstract}
To overcome current serious energy and environmental issues, photocatalytic water splitting holds great promise because it requires only solar energy as an energy input to produce hydrogen. Two-dimensional (2D) semiconductors and heterostructures possess several inherent advantages which are more suitable for boosting solar energy than their bulk counterparts. In this work, by performing high-throughput first-principles calculations combined with a semiempirical van der Waals dispersion correction, we first provided the periodic table of band alignment type for van der Waals heterostructures when packing any two of the 260 semiconductor monolayers obtained from our 2D semiconductor database (2DSdb) (\url{https://materialsdb.cn/2dsdb/index.html}). Based on the rules of thumb for photocatalytic water splitting, we have further screened dozens of potential semiconductors and thousands of heterostructures which are promising for photocatalytic water splitting. The resulting database would provide a useful guidance for experimentalists to design suitable 2D vdWHs and photocatalysts according to desired applications.
\end{abstract}
\maketitle

\section{Introduction}
Exploiting photocatalytic water splitting (PWS) for hydrogen production offers a potential solution to the energy crisis caused by the approaching exhaustion of fossil fuels and the serious environmental problems.\cite{Turner2004} The discovery of PWS on TiO$_2$ electrode by Fujishima and Honda in 1972 has opened a new epoch in harnessing solar energy.\cite{FUJISHIMA1972} Many potential semiconductor photocatalysts have been extensively investigated and various photocatalysis,\cite{Wang2014a} including WO$_3$,\cite{Chen2008} Bi$_2$WO$_6$,\cite{Zhang2007} Bi$_2$O$_3$,\cite{Zhang2006} C$_3$N$_4$,\cite{Wang2009} and CdS,\cite{Bao2008} have been developed. However, the practical application of PWS for hydrogen production is limited by several technical issues such as the inability to utilize visible light, insufficient quantum efficiency, fast backward reaction, and poor activation of catalysts.\cite{Liu2011,Bak2002,Ni2007,Wang2020} In bulk materials, photoexcited carriers must move to the surfaces of photocatalysts to participate in the redox reaction with the absorbed water molecules. Due to the long and complex migration path, a large number of photoinduced electrons and holes may recombine, resulting in poor photocatalytic efficiency.\cite{Wang2020} Continuous search of efficient and environment-friendly photocatalysts for hydrogen production is therefore very urgent.

Two-dimensional (2D) semiconductors have many advantages which can be used to improve the photocatalytic efficiency for water splitting as discussed in several reviews and the references therein.\cite{Singh2015,Zhang2018,Faraji2019,Wang2022,Fu2022,Jakhar2022} Compared with their bulk counterparts, 2D materials can provide a more efficient redox reaction due to the abundant reactive sites, improved electron-hole separation, fast mobility of charge carriers, short diffusion length for photo-generated electrons and holes, which causes lower electron-hole recombination rate and better quantum efficiency.\cite{Singh2015,Fu2022} Thus, the search and design of 2D photocatalysts for water splitting has been a hot topic in both experimental and theoretical studies. \cite{Singh2015,Fu2022} Currently, a variety of 2D materials have been verified to be promising photocatalysts for water splitting verified by experiments or first-principles calculations, including MoS$_2$,\cite{Li2013} g-C$_3$N$_4$,\cite{Liu2016a} SnS$_2$,\cite{Sun2012} FePS$_3$,\cite{Cheng2018} single layer group-III monochalcogenides,\cite{Zhuang2013} Transition metal phosphorus.\cite{Shifa2018} Owing to the quantum confinement effect, 2D photocatalysts can be further modulated through band gap engineering by varying the chemical and/or physical conditions, including layer thickness, stacking order, doping, defect engineering, strain, external electric field, etc.\cite{Li2017,Ganguly2019,Fu2022} 

However, 2D materials also face some problems while serving as photocatalysts.\cite{Li2017,Su2018} For example, photogenerated carriers accumulated in monolayers are easy to recombine. Moreover, some 2D semiconductors are not stable in air or aqueous solution, leading to the decrease of the photocatalytic activity. In parallel with the efforts on synthesis of new 2D monolayer photocatalysts, another strategy has been gaining strength over the past few years. By stacking together different 2D materials on top of each other, various artificial van der Waals heterostructures (vdWHs) can be formed.\cite{Geim2013,Novoselov2016} Compared with the conventional bulk semiconductor-based heterostructures, vdWHs do not demand crystal lattice matching, allowing to build artificial vdWHs with desired functionalities by picking and stacking atomic layers of arbitrary compositions. The vdWHs not only preserve the excellent properties of the original single layers due to the weak vdW interaction, but also bear additional features. The construction of vdWHs is an effective method to further improve the photocatalytic performance as discussed in several reviews.\cite{Singh2015,Zhang2018,Shifa2019,Wang2022,Jakhar2022} Especially, in type-II heterojunctions, the lowest energy states of holes and electrons are on different sides of the heterojunctions, 
which ensures the effective separation of photogenerated electrons and holes, thereby suppressing the recombination of photogenerated electrons and holes.\cite{Wang2013,Low2017}

Up to now, the electronic and optical properties and photocatalytic applications of many vdWHs have been predicted through first-principles calculations, providing strong support for experimental synthesis and commercial applications.\cite{Wang2013,Chen2022} In our previous study, we performed high-throughput first-principles calculations to build a 2D semiconductor database (2DSdb) which including more than 260 2D semiconductors.\cite{Wang2018a} In spite of the rapid development of high-throughput computational 2D crystals databases publicly also available at present, such as MC2D,\cite{Mounet2018} C2DB\cite{Haastrup2018}, 2DMatPedia\cite{Zhou2019} and JARVIS-DFT\cite{Choudhary2017}, systematic high-throughput predictions of band alignment in vdWHs are still rather incomplete and investigations of 2D vdWHs applied in photocatalysis are strongly called for. 

In this work, combining the high-throughput first-principles calculations with a semiempirical vdW dispersion correction, the periodic table of heterostructure types including around 34000 possible vdWHs is obtained based on Anderson's rule. Furthermore, dozens of potential semiconductors and thousands of vdWHs for PWS applications have been further screened based on the rules of thumb for PWS.  The remainder of this paper is organized as follows. In Sec. II, methodology and computational details are described. The details of screening criteria are discussed in Sec. III. Sec. IV presents high-throughput computational screening of 2d semiconductors and heterostructures potential for photocatalytic water splitting. Finally, a short summary is given in Sec. V.

\section{Methodology}
\subsection{Density Functional Calculations}

Our total energy calculations were performed using the Vienna Ab initio Simulation Package (VASP).\cite{Kresse1996, Kresse1996a} The electron-ion interaction was described using projector augmented wave (PAW) method \cite{PAW, Kresse1999} and the exchange and correlation (XC) were treated with GGA in the Perdew Burke Ernzerhof (PBE) form\cite{Perdew1996}. Part of electronic structure calculations were also performed using the standard screening parameter of Heyd-Scuseria-Ernzerhof (HSE06) hybrid functional, \cite{Becke1993,Heyd2003,Perdew1996a,Paier2006,Krukau2006,Marsman2008}  upon the PBE-calculated equilibrium geometries. A cutoff energy of 400 eV was adopted for the plane wave basis set, which yields total energy convergence better than 1 meV/atom. In addition, the non-bonding vdW interaction is incorporated by employing a semi-empirical correction scheme of Grimme's DFT-D2 method in this study, which has been successful in describing the geometries of various layered materials.\cite{Grimme2006, Bucko2010}  In the slab model of 2D systems, periodic slabs were separated by a vacuum layer of 20 {\AA} in \emph{z} direction to avoid mirror interactions. 
The Brillouin zone was sampled by the  \emph{k}-point mesh following the Monkhorst-Pack scheme,\cite{Monkhorst1976}  with a reciprocal space resolution of 2$\pi$$\times$0.03 {\AA}$^{-1}$. On geometry optimization, both the shapes and internal structural parameters of pristine unit-cells were fully relaxed until the residual force on each atom is less than 0.01 eV/{\AA}.  To screen the novel 2D hotocatalysts, we used the VASPKIT package\cite{vaspkit} as a high-throughput interface to pre-process the input files and post-process the data obtained by using VASP code.

\section{Results and Discussions} 
\subsection{2D van der Waals Heterojunctions}

We begin our discussion by determining the band alignments of vdWHs when packing any two of the 260 semiconductors obtained from our previous study.\cite{Wang2018a} According to the alignments of the CBM and VBM in the constituent layers, vdWHs can be classified into three types: type I (straddling gap), type II (staggered gap), or type III (broken gap), as illustrated in Fig. \ref{align_type}(a), respectively. In type I heterojunctions, both VBM and CBM of two independent component semiconductors are located at the same side of the heterointerface. This is beneficial for spatially confining electrons and holes so that efficient recombination can be achievable, rendering them potential applications in optoelectronic devices such as lightemitting diodes (LEDs).\cite{Withers2015} In type II heterojunctions, the CBM and VBM are located in different components with electrons accumulating in the layer with the lower CBM and holes accumulating in the other layer with the higher VBM. Different from type I band alignment, the separation of electrons and holes on different layers can increase carrier lifetime, which is desirable for  photocatalysis and unipolar electronic device applications including  photovoltaics and photodetection applications\cite{Hong2014,Furchi2014,Lin2015,Massicotte2016,Deng2016,Furchi2014} In type III heterojunctions, the VBM of one semiconductor is higher than the CBM of the other, making the whole system  overall heterojunction metallic. Such a property could have great potential in tunnel field-effect transistors and wavelength photodetectors.\cite{Yan2015,Shim2016}

\begin{figure}[htbp]
\centering
\includegraphics[scale=0.42]{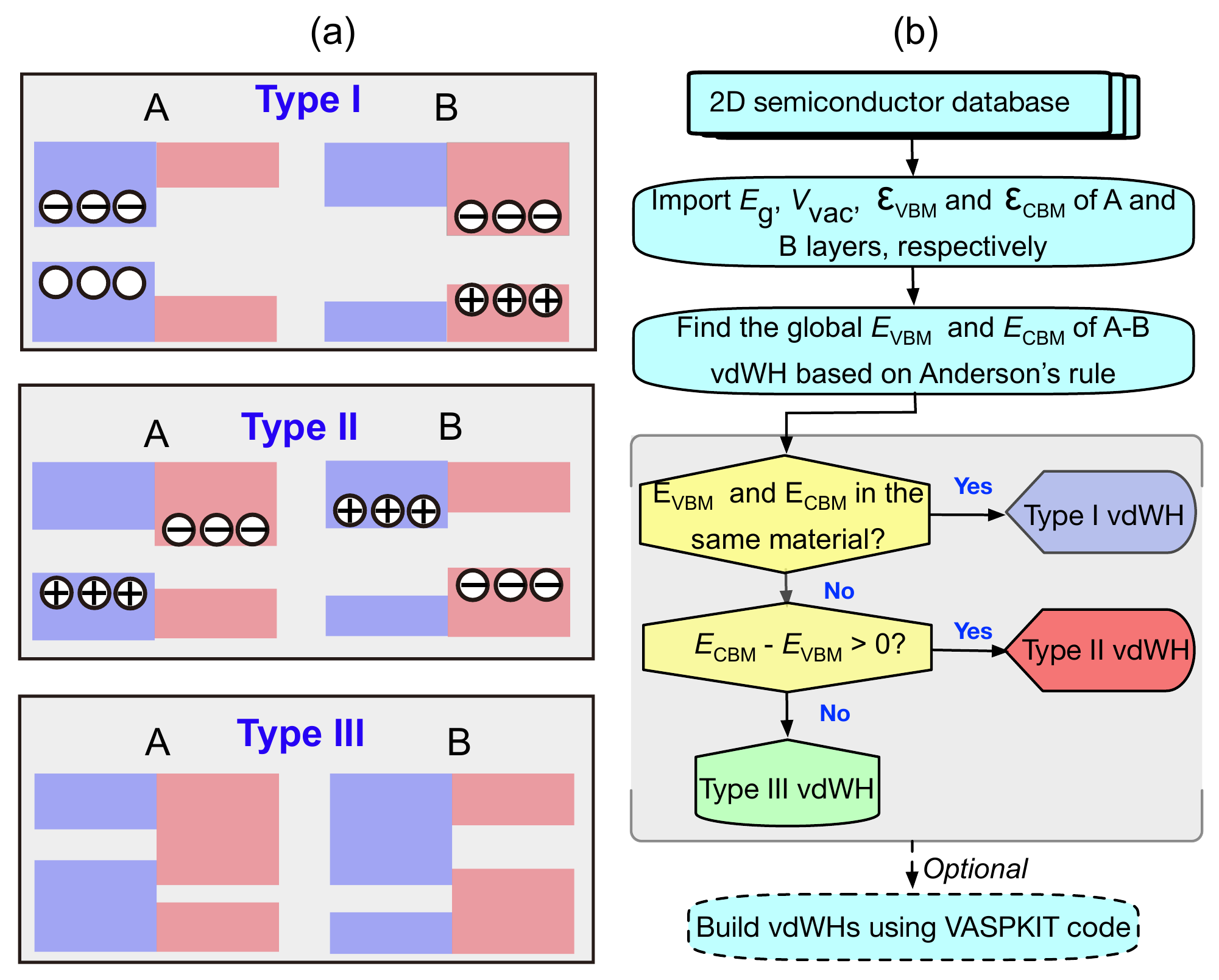}
\caption{\label{align_type}(Color online) (a) Schematic illustrations  of the three types of semiconductor heterojunctions based on their energy band alignments: type I (straddling gap), type II (staggered gap), and type III (broken gap) heterojunctions. (b) flowchart for the computational design of vdWHs.}
\end{figure}

The high-throughput design of vdWHs  has gained significant attention because vdWHs have unique physical properties and potential applications mentioned above. Rasmussen \emph{et al}. theoretically predicted the band alignments of 51 semiconducting TMDs and TMOs monolayers using GW0 calculations.\cite{Rasmussen2015} Latterly, {\"a}z{\c c}elik \emph{et al}. established a periodic table of band alignments cover about 900 vdWHs by performing hybrid functional calculations.\cite{Tony2016} Nevertheless, these studies focused only on groups IV, III-V and V elemental and/or compound monolayers, TMDs and transition-metal trichalcogenides (TMTs). To have a thorough search, we have extended the HSE06 calculated periodic table of heterostructure types formed by any two of the 200 screened semiconductors. The flowchart for the computational design of vdWHs is illustrated in Fig. \ref{align_type}(b). To determine the band alignment type  when A and B monolayers were stacked together, we first compared the absolute positions of VBM and CBM by aligning the vacuum level of two composed monolayers to 0 eV based on Anderson's rule.\cite{Anderson1960}  The band gap of vdWH was then estimated by the difference between the lower CBM energy and the higher VBM energy of two monolayers. 
The resulting vdWH is type I if both the higher VBM and  lower CBM are located at the same layer. Otherwise, it belongs to type II or type III. Next the II and III can be  separated by the band gap of vdWH being larger than zero or not.

\begin{figure*}[htbp]
\centering
\includegraphics[scale=0.66]{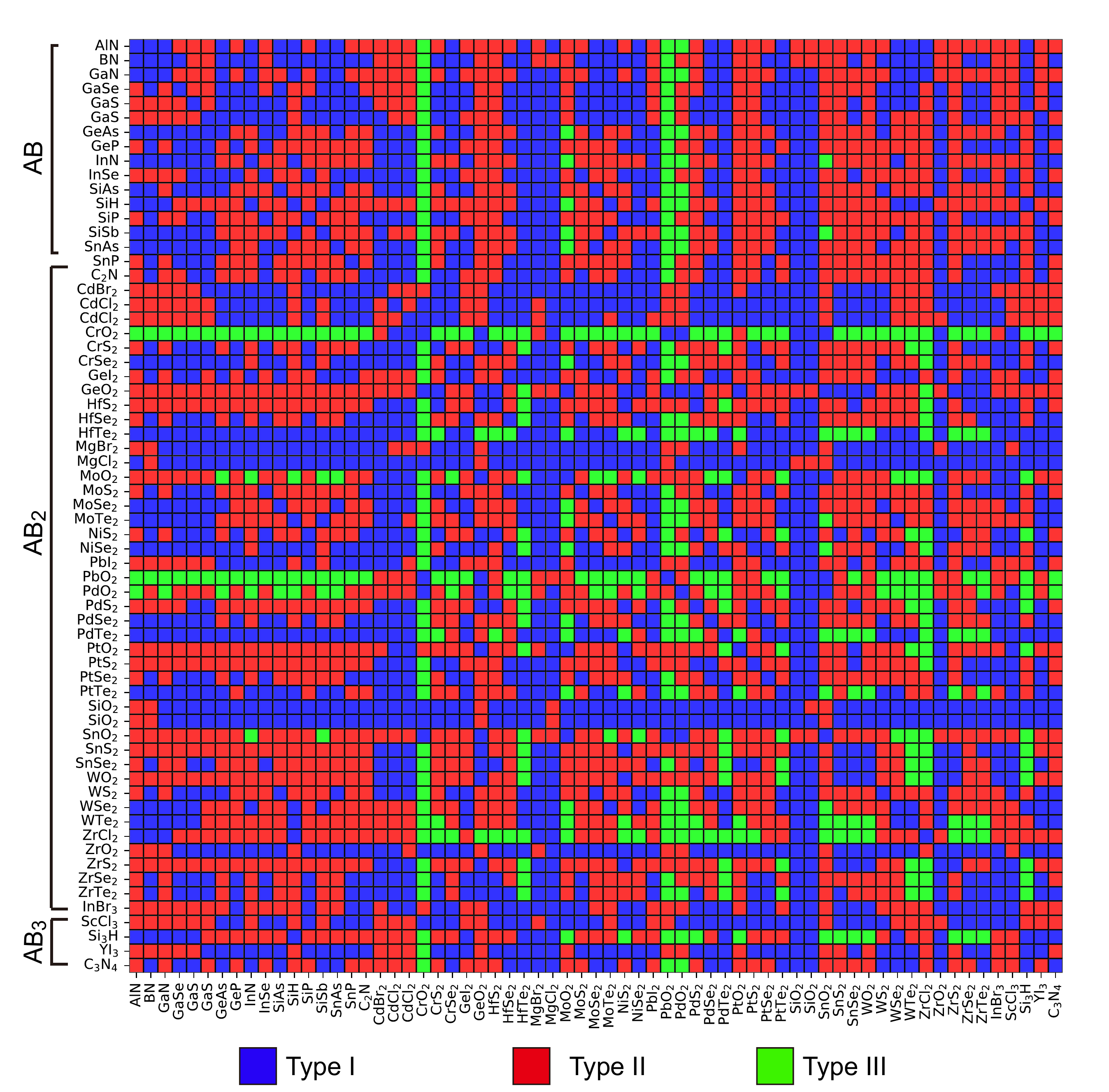}
\caption{\label{ptable_hex}(Color online) HSE06 calculated periodic table of hexagonal vdWHs. Type I (44.5\%), II (44.9\%), and III (10.6\%) band alignments are represented by blue, red, and green boxes, respectively.}
\end{figure*}

Considering that the dimension table of complete vdWHs type is very huge (around 34000), we only list the periodic table of vdWHs for about 2000 hexagonal systems in Fig. \ref{ptable_hex}. 
Our results show that the hexagonal vdWHs are dominated by type II (44.6\%) and type I (43.8\%), and the least abundant are type III (11.6 \%). In contrast, the complete periodic table provided in the Supplemental Material shows that almost half of possible vdWHs are type I (44.0\%), followed by type II (40.6\%) and type III (15.4\%). One can find that  the vdWHs composed of light elements are dominated by type I and II (near the upper left corner); while the vdWHs composed of heavy elements tends to form type III heterojunctions.  This is mainly because the 2D semiconductors composed of heavy elements have higher ionization energy than those composed of light ones. Overall, the agreement between our calculated results and the corresponding experimental values is very good.  For examples, ultrafast charge transfer was observed in MoS$_2$/WS$_2$ type II heterojunction,\cite{Hong2014}  and moir{\'e}-trapped valley excitons was observed in MoSe$_2$/WSe$_2$ type II heterojunction,\cite{Seyler2019}

To validate the band-alignment description with the predictions from Anderson's rule, three representative heterojunctions with relatively small lattice mismatch, ZrO$_2$/MoSe$_2$, WSe$_2$/MoSe$_2$ and NiS$_2$/ZrCl$_2$ were chosen to revisit their band alignments by performing HSE06 calculations. Their layer-resolved band structures are shown in Fig. \ref{align_type}. One can find that the band alignments predicted using Anderson's rule is similar to the DFT results.  As expected, no significant interlayer hybridization between the two components of vdWHs is found. 

It needs to be emphasized that the Anderson's rule-based approach is a rough estimation. It may work only qualitatively as the interlayer coupling plays an important role in determining the band structures of 2D materials.\cite{Liu2016,Novoselov2016,Shi2018,Wu2022} Part of the interlayer coupling arises from the electronegativity difference between the two layers, which might cause band shifts and band gap variations in vdWHs. The magnitude of interlayer coupling dependent on layer-thickness, interlayer distance, stacking order, interlayer twist angle, constituent elements, the symmetry of two components, $etc$.  Furthermore, for the lattice-mismatched vdWHs, the internal stress between the constituent layers is more likely to modify their band structures in addition to direct interlayer coupling.  Thus, it is expected that the vdWH type can vary due to the interlayer coupling and internal stress if the difference of absolute band extrema of two components is not significant. Nevertheless, the effect of interlayer coupling on the electronic structures of vdWHs is rather complicated and beyond the scope of the present work. 

\begin{figure*}[htbp]
\centering
\includegraphics[scale=0.5]{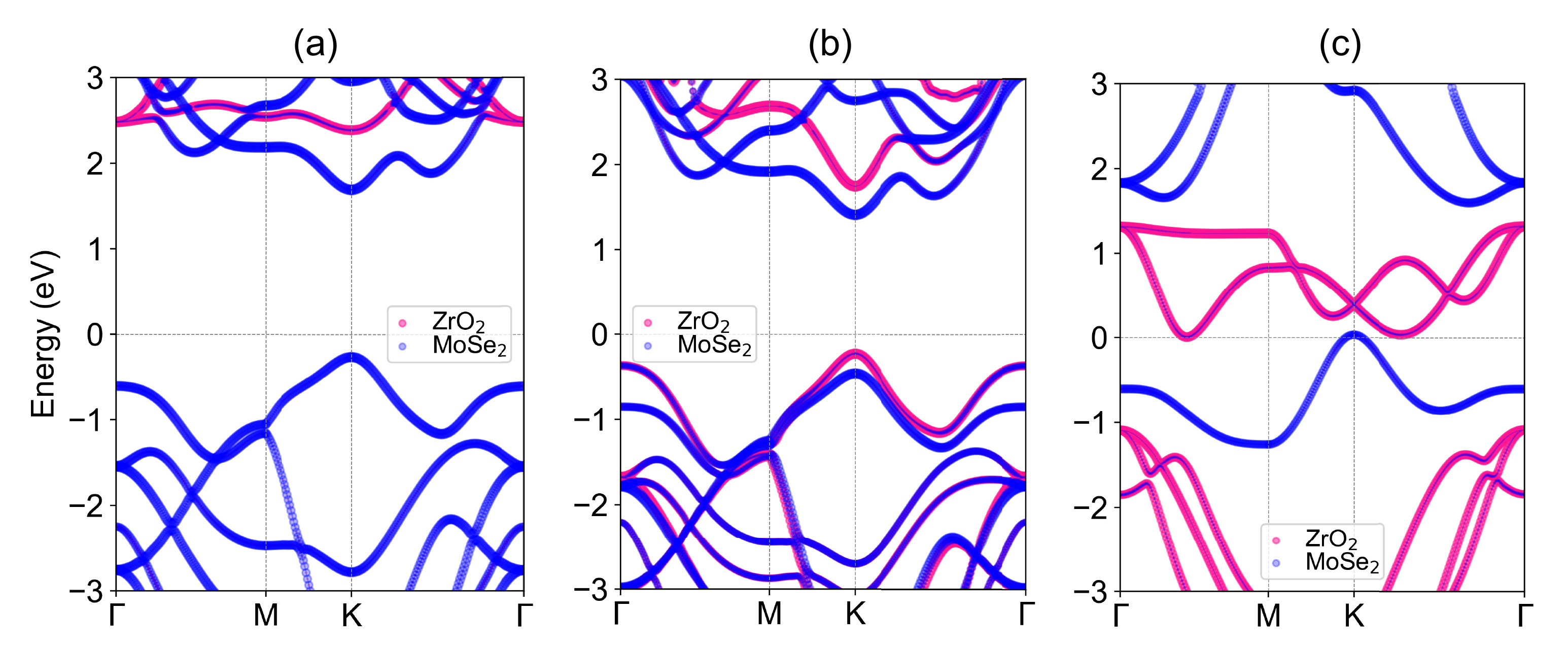}
\caption{\label{align_type}(Color online) Layer-resolved band structures of (a) ZrO$_2$/MoSe$_2$ with type I junction, (b) WSe$_2$/MoSe$_2$ with type II junction and (c) NiS$_2$/ZrCl$_2$ with type III junction, respectively. The Fermi level is shown with black dashed line.}
\end{figure*}

\subsection{2D Semiconductors and Heterojunctions Photocatalysts}
To be a promising candidate  for water splitting, a semiconductor should meet three basic requirements:\cite{Singh2015,Chakrapani2007} (i) being chemically stable and insoluble in water; (ii) band gap being larger than the free energy of water splitting of 1.23 eV and  smaller than 3 eV to enhance solar absorption; and (iii) band edge position crossing the redox potentials of water,\cite{Walter2010} \emph{i.e.,}, the CBM being higher than the reduction potential of H$^+$/H$_2$ (-4.44 eV at pH = 0) and the VBM being lower than the oxidation potential (O$_2$/H$_2$) (-5.67 eV at pH = 0). Moreover, the redox potentials is influenced by the pH value in the water splitting reaction. Specifically, the pH-dependent reduction potential for H$^+$/H$_2$  and oxidation potential for O$_2$/H$_2$ are $ E_{\mathrm{H}^{+}}^{\mathrm{red}} / \mathrm{H}_{2}= -4.44 +\mathrm{pH} \times 0.059$  eV and $E_{\mathrm{O}_{2} / \mathrm{H}_{2} \mathrm{O}}^{\mathrm{ox}}=-5.67 +\mathrm{pH} \times 0.059$ eV respectively.

\begin{figure*}[htbp]
\centering
\includegraphics[scale=0.9]{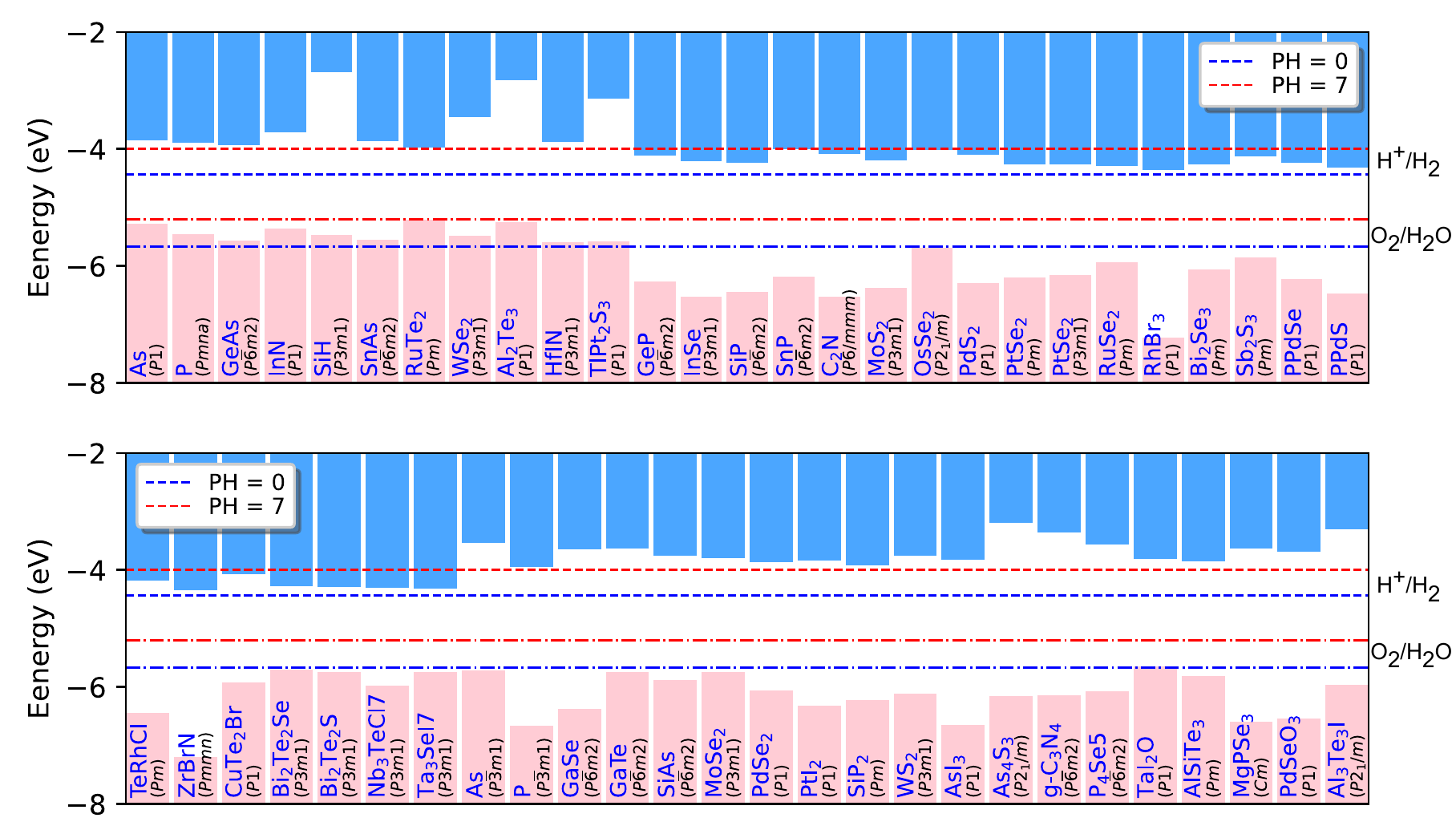}
\caption{\label{photocatalysis}(Color online) HSE06 calculated band edge positions of 2D candidates relative to the vacuum level for PWS. 
The position of the reduction level for H$^+$ to H$_2$ and the oxidation potential of H$_2$ to O$_2$ are indicated by the dotted and dashed lines, respectively.}
\end{figure*}

On the basis of above-mentioned criteria (ii) and (iii), we have extended the photocatalysis screening procedure to our 2D semiconductor database. We obtained 53  kinds of monolayers possessing band edge positions which meet the requirement for PWS, as shown in Fig. \ref{photocatalysis}. These results include the extensively studied 2D semiconductors. For examples, recent experimental or computational studies have revealed that graphene-like $g$-C$_3$N$_4$,\cite{Wang2009} C$_2$N,\cite{Mahmood2015} MoS$_2$,\cite{JaramilloThomas2007} BP,\cite{Hu2017} PdSeO$_3$,\cite{Qiao2018} SiP$_2$,\cite{Yu2021}  Bi$_2$Te$_2$S and Bi$_2$Te$_2$Se\cite{Wang2018} are potential candidates for photocatalytic water splitting. Besides the previously reported monolayer materials, we have also screened around 40 novel monolayer semiconductors with good structural stability and proper band edge positions.  Among them, there are 17 potential candidates being applicable to a wide range of pH value from 0 to 7. However, it should be emphasized here that the presence of appropriate band edge positions cannot guarantee an effective photocatalyst for water-splitting. The chemical stability of these candidates in exposure to water and air were not considered in our high-throughput screening calculations. Furthermore,  the fast recombination of photoinduced carriers is the main limitations for photocatalytic water-splitting since the photo-excited carriers in such thin layer can quickly recombine. Zhang and co-workers reported that the semiconductors with indirect band gap character can reduce the possibility of recombination of photogenerated electrons and holes.\cite{Zhang2014}

Aside from monolayer 2D semiconductors, the type-II vdWHs,  have also been developed as a new avenue to design high-performance photocatalysts for water splitting. Previous experimental and theoretical studies focus on a limited number of vdWHs, such as C$_2$N/MoS$_2$,\cite{Guan2017,Kumar2018}C$_2$N/WS$_2$, \cite{Kumar2018} $g$-C$_3$N$_4$/C$_2$N,\cite{Wang2016} $g$-C$_3$N$_4$/MoS$_2$,\cite{Wang2014} $g$-C$_3$N$_4$/BP,\cite{Ran2018} and InSe/$g$-C$_3$N$_4$.\cite{He2019}
Zhang \emph{et al}. screened 44 kinds of potential type-II heterojunctions for water splitting under the constraint of suitable band edge positions and similar lattice parameters.\cite{Zhang2018} Here from more than 30000 possible vdWHs aforementioned, we have screened 1994 (pH = 0), 2872 (pH = 7) and 1865 (pH = 0$\sim$7) type-II vdWHs which have potential for water splitting photocatalysts. Detailed information are given in the Supplemental Material, respectively.  It is expected that our results will serve as a guide to improve the photocatalytic performance of 2D materials.

\section{Summary}
In conclusion, we have identified the types of band alignment for $\sim$34000 van der Waals heterostructures. Based on the rules of thumb for photocatalytic water splitting, we have further screened 53 monolayer semiconductors, 1994 (pH = 0), 2872 (pH = 7) and 1865 (pH = 0$\sim$7) vdWHs which have potential for water splitting photocatalysts. We hope that our computational screening database could stimulate further exploration of heterostructures in nanoscale devices, photocatalysis applications, and other important applications.

\section{Acknowledgement}
This work was supported by National Natural Science Foundation of China (Grant No. 62174136). In addition, this work was also partly supported by the Natural Science Basic Research Program of Shaanxi (Program Nos. 2022JQ-063 and 2021JQ-464), The Natural Science Basic Research Plan of Shaanxi Province (Grant No. 2021JZ-48), The Scientific Research Program Funded by Shaanxi Provincial Education Department (Grant Nos. 21JP088 and 22JP058) and The Youth Innovation Team of Shaanxi Universities. The calculations were performed on the MASAMUNE-IMR supercomputer at the Institute for Materials Research of Tohoku University, Japan (Project No.2112SC0503). J.N. were supported by Innovative Science and Technology Initiative for Security Grant Number JPJ004596, ATLA, Japan.

\nocite{*}
\bibliographystyle{aipnum4-1}
%

\end{document}